\documentstyle[12pt]{article}
\begin{document}
\baselineskip=24pt
\begin{center}
{{\Large Symmetrized DMRG studies of the properties of Low-lying states of 
Poly-para-phenylene (PPP) system} \footnote{Contribution no. 1254 from
Soild State and Structural Chemistry Unit}}
\end{center}
\begin{center}
{\bf Y. Anusooya$^{2}$, Swapan K. Pati$^{2}$ and S. Ramasesha$^{2,3}$}
\end{center}
\centerline{$^2$Solid State and Structural Chemistry Unit}
\centerline{ Indian Institute of Science, Bangalore 560012, India}
\centerline{$^3$Jawharlal Nehru Center for Advanced Scientific Research}
\centerline{ Jakkur Campus, Bangalore 560064, India}

\begin{center}
{\bf Abstract}
\end{center}
We report the symmetrized density matrix renormalization group (DMRG) study
of neutral and doped oligomers of PPP system within an extended Hubbard
model. Model parameters are determined by comparing the existing results
for interacting small system. We compute a number of properties in the
ground state as well as in the one-photon, two-photon and triplet states
to completely characterize these states. Based on the relative positions 
of the one-photon and two-photon states in long oligomers, we suggest that 
invoking interchain interactions is essential for explaining strong 
fluorescence in this system. Bond-order studies show that the
lowest two-photon state corresponds to a localized excitation while
one-photon and triplet excitations are extended in nature. The bipolaronic
state shows clear evidence for charge separation and disproportionation 
into two polarons. We find that the extended nature of one-photon and 
triplet states of neutral system are very similar to those of the 
bipolaronic ground states.

\pagebreak

\section{Introduction}
Electronic structure of conjugated organic polymers has been a field
of enduring interest from both experimental and theoretical stand points.
The insulating behaviour of these systems coupled with different topologies
in which they exist has thrown open the possibility of a vast variety of 
qualitatively 
different excited states. In systems with degenerate ground states, we could
have solitonic excitations besides the polaronic or bipolaronic excitations
of doped systems\cite{bred}. These excitations are primarily a consequence of
electron-lattice interactions. The insulating nature of the polymers implies
fairly long-range electron-electron interactions. In neutral systems this
leads to the possibility of excitonic, biexcitonic or indeed exciton-string
like excitations\cite{sumi}. Besides, these low-lying excitations also 
have spin
degrees of freedom leading to their classification as singlets, triplets, etc.
The excitations can further be labelled by the symmetry label of the 
irreducible representation according to which a given state transforms. 
The inter-play of electron-lattice and electron-electron interactions leads
to complicated yet very interesting and important physical situations. 

The polymers that have attracted much attention in recent years are the
poly-para-vinylene (PPV) and poly-para-phenylene (PPP) systems. Both these
polymers have the {\it para} position of the phenyl rings in conjugation.
It was earlier suggested by Soos et.al\cite{soos} that such a system can
be mapped onto a strongly dimerized polyacetylene chain and a set of 
isolated ethylenic units, in the noninteracting limit. Due to the strong
dimerization introduced by the transformation, it is also suggested that 
the electronic excitations in the presence of correlations would be
more like that of a noninteracting system. This has very strong
implications for the fluorescence properties of these polymers. In the
noninteracting picture, the lowest siglet excitation is dipole
allowed and in these cases we should expect strong fluorescence since
from Kasha's rule\cite{kasha} it is known that fluorescence occurs
from lowest singlet state. However, polymers such as polyacetylene (PA)
the lowest excited state is a dipole forbidden state\cite{zgs} since
the dominant effect of electron correlation is not suppressed by the
moderate dimerization in these systems. The lowest excited singlet
state being dipole forbidden in the ideal limit, we should expect
weak fluorescence. This brings into focus the symmetry of the 
low-lying excitations in these polymers. 

Electronic energy levels of doped polymers are also of considerable
interest, since polymers are envisaged as active components in
light emitting diodes (LED). The polymers which have been identified
as luminescent semiconductors thus far include PPP, Poly-para-phenylene-
vinylene(PPV), polythiophene
and their derivatives\cite{frin}. To tailor the properties
of polymers for such applications, it is necessary to understand the
doped states in these systems that lie below the conduction band. Many 
different experimental techniques have been employed to study 
these ``in-gap'' states.
They are photoinduced absorption (PA)\cite{abe,wei,hsu}, 
photoconductivity (PC) measurements\cite{lee},
electroluminescence (EL)\cite{swan} and photoluminescence 
(PL)\cite{sumi,hero}, to name a few. 
These studies  have often been carried out using pump-probe techniques to 
elicit the dynamics and decay of the electronic excited states\cite{hall}.

Theoretical studies of  conjugated polymers have been approached from 
two different stand points. In one approach, quantum cell models such as 
the Pariser-Parr-Pople model with static electron-lattice interactions
have been studied employing mainly exact diagonalization techniques for
small system sizes. The study of large system sizes have not been quite
satisfactory in any of the approaches to these models, until recently. 
However, the advantage of model Hamiltonian studies is that they allow 
exploring the parameter space
freely to elicit information in different limits of the models.

The second approach has been pioneered by quantum chemists who employ
empirical Hamiltonians such as those obtained from intermediate neglect
of differential overlap (INDO) approximation\cite{ind}. Using such 
Hamiltonians,
oligomers of several units are studied within the framework of a 
Hartree-Fock
solution followed by limited conifguration interaction (CI) calculations
\cite{jlb}.
The issues that are addressed in these studies revolve around the
optimal geometries of molecules, effects of derivatization of monomers and
to a limited extent the spectroscopic characteristics of the 
system\cite{bred1}. The
main advantage of these calculations over the quantum lattice models is
that we can study systems with rather large unit cells within a realistic
approach but rather poorly (when only single or double CI are performed) for
excited state properties, although, the ground state properties such as 
geometries are known to be predicted quite accurately by these 
methods.

In recent years, the density matrix renormalization group (DMRG) method
has been successful in providing accurate solutions to model Hamiltonians
with short-range interactions in low-dimensions\cite{white}. Inclusion of 
symmetry in
the DMRG procedure has further allowed access to excited states\cite{srskp}. 
The minimal
model that can support an excitonic state is the ``U-V'' model. 
Reliable solutions of quantum cell models for excited states can 
clarify the physics of various kinds of ``in-gap'' excitations. 
In this paper, we have placed emphasis on the nature of the excitations and
relative positions of the excitations rather than on comparison between
theory and experiment. The absolute comparisons are made difficult by the 
substituent groups in actual system and furthermore, the ``U-V'' model
parametrization is not likely to include the effective long-range
interactions that are present in the actual system. We employ
the DMRG technique to study the low-lying excitations in PPP system
within the ``U-V'' model in the neutral state and with both electron and 
hole doping. In Fig. (1) is shown the schematic diagram of the left-half
of the PPP system we have studied. We follow the labelling of
the sites and bonds as given in Figs. (1a) and (1b) respectively. In the next 
section, we introduce modification to the 
usual DMRG procedure\cite{white}. This is followed 
by a section on results and discussion. The last section summarizes the 
contents of the paper.

\section{Symmetrized DMRG method for PPP system}
All the carbon sites in the PPP system are taken to be identical
and this idealized situation leads to alternancy symmetry in the system.
Besides, this system possess reflection or $C_2$ symmetry 
about the central bond of the polymer in the planer geometry. The ``U-V'' 
Hamiltonian which is 
used to model the electronic states is spin independent and hence conserves
both $S^z_{total}$ and $S_{total}$. We employ only parity which bifurcates
the Hilbert space into space of even and odd total spins\cite{affleck}. 
These three
symmetries (alternancy, $C_2$ or reflection and parity) commute 
with each other resulting in an Abelian group consisting of eight elements
and the eight irreducible representations that exist can be labelled by the 
indices A/B, +/- and o/e corresponding to reflection, alternancy and parity
symmetries. The ground state of the singlet for the neutral polymer exists
in the $^eA^+$ subspace corresponding to the even spin, ``covalent'' - A
subspace. The first dipole allowed transition is from the ground state to 
the lowest energy state in $^eB^-$ subspace corresponding to even spin 
``ionic'' $B$ space. The lowest triplet state is to be found in $^oB^+$ 
subspace. Doping the system leads to breaking of alternancy symmetry
and doped states are characterized  by total spin, z-component  of total 
spin and A/B label. In the symmetrized DMRG procedure, we have taken care
to carry out the calculation of the lowest state in each subspace, before
indentifying the ground state and the excitations.

The oligomers of PPP are constructed starting from a four site
initial system.  Fig. (2) shows the manner in which the sites are 
added on the left and right parts
of the system to build up the PPP oligomers. The
system evolves from the external carbon atoms of the
chosen oligomers (Fig.2) and the carbon sites are progressively 
introduced  in the
middle. This procedure is very accurate compared to the procedure
that builds the 
oligomer starting from the middle of the total system. This is because,
in the latter, we have to rebuild the central bonds each time and the 
repeated renormalization
of the operators at this site reduces the accuracy of the procedure.
Besides, the inside-out procedure does not introduce interactions between
the new sites and thus the density matrix of the half-blocks are not
good representatives of the density matrix of the half-block of the
complete oligomer. The order in which the sites are introduced also
avoids interaction between sites that are introduced several iterations apart.
In our scheme a newly added site interacts atmost with a site introduced
two  iterations previously. This procedure should therefore be as accurate
as a quasi-one-dimensional system with nearest and next nearest neighbour
interactions or the ladder systems. 
This process of construction retains the symmetries of the full Hamiltonian 
at every stage and hence we can target the lowest state in the desired 
subspace at each iteration. The symmetrized DMRG procedure is implemented
along the lines described in an earlier paper\cite{srskp}. 

The geometry of phenyl groups in PPP are taken to be of equal bond lengths
(corresponding to the value of $1.397 \AA$ in benzene) with uniform transfer
integrals $t_0=2.4$eV. The inter phenyl bonds have a transfer integral of 
$2.077$eV corresponding to a bond length of $1.508\AA$, the system is
however taken to be planar; i.e,  the dihedral angle between successive
phenyl rings is taken to be zero.  The $U$ and $V$ parameters
are chosen such as to reproduce the 
Pariser-Parr-Pople values with standard parameters for biphenyl for the 
optical gap. These correspond to $U=11.26$eV and $V=4.5$eV. In Table (1)
we compare the exact excitation energies for biphenyl with standard
Pariser-Parr-Pople parameters with those of the ``U-V'' model. We also 
compare, in this table, the excitation energies from the DMRG procedure 
for the  ``U-V'' model. 

Table (1) shows that the $U$ and $V$ values chosen in the ``U-V''
model are reasonable. Besides, it also shows that the DMRG method is accurate 
for small systems. For large systems, confidence in DMRG method can 
be gained by comparing
the DMRG results for the noninteracting system ($U=V=0$) with exact
noninteracting results.
When the interactions are turned-off, the models are exactly solvable by 
one-electron methods for the largest system sizes we have addressed. 
 We compare the results from the
one-electron method with the DMRG results for the non-interacting models
of PPP for various dopings for the largest system we have studied
for a DMRG cut-off of $m=120$. The quantities we compare are energies 
(Table (2)) and bond orders for the ground states as well as doped states
 of the longest oligomer of PPP we have studied for different dopings
(Figs. (3-5)). We find the comparison between DMRG and exact results
 to be excellent in the noninteracting limit, where DMRG has the least 
accuracy\cite{whitescal}. Because of the desired accuracy of the 
infinite DMRG procedure we have chosen not to carry-out the finite DMRG 
procedure which is very compute intensive for the cut-off $m$ that we have 
imposed\cite{white,anu}.

\section{Results and Discussion}

We have carried out ``U-V'' model calculations of PPP systems with upto
eight phenyl rings, for neutral as well as
hole and electron doped cases. We have obtained energies and bond-orders
of ground state and also for a few low-lying excited states. Besides, we have 
computed charge and spin densities, charge-charge correlation functions.

We have plotted the change in energy in the ground state for the PPP
system in going from $(n-1)$ ring system to an $n$ ring system
 in Fig. (6). We find that, there is a clear $even-odd$ pattern
in the ground state energy per ring. Addition of a phenyl ring
to an even ring PPP oligomer stabilizes the polymer by a slightly
greater extent than when a phenyl ring is added to an odd ring system.
This difference should vanish when we go to a system with very large
number of rings. The above result shows that the finite size effect
is persistent even for an 8-ring oligomer, although the magnitude
of the effect itself is rather small. The difference in energy per ring
when a ring is added to an even system as opposed to a ring added
to an odd system is $0.024$ eV, while the average energy per ring is
$-11.428$ eV.

In Fig. (7), we present excitation energies for the one-photon,
two-photon and lowest triplet excitation for the oligomers of PPP.
The optical (one-photon) gap shows a sharp drop in going from the
dimer to the oligomers and for the long oligomers lies between
the values for the 8-ring system  ($4.667$eV) and  for the
7-ring system ($4.45$eV). The two-photon gap is slightly more than 
$1$eV below the optical gap. The two-photon gap also shows finite 
size effects although it is less pronounced than that in the 
one-photon  gap. The 
singlet-triplet gap is at an energy of $\sim 2.4$eV which is
$2$eV below the optical gap and this gap shows the least 
finite size effects amongst the excitation gaps we have studied.

We find from our studies that the one-photon gap is always above
than the two-photon gap in the PPP system. This is contrary to the
excitation from the mapping of the system to a strongly dimerized
chain in the noninteracting limit\cite{soos}. It appears that in the
presence of electron correlations, the mapping is not accurate enough
to conclude about the relative positions of the excited states.
If we introduce nonplanarity, through nonvanishing dihedral angles
between successive rings, the classification of states by the labels
$A$ and $B$ is lost due to the broken $C_{2}$ symmetry of the system.
The optical excitation would then correspond to the transition from the
ground state to the lowest state in the ionic space. However, we know
from earlier studies on biphenyl\cite{bhaba} that the one-photon state 
in this case is also derived from the $B$ space and has a higher energy gap
than in the planer system. To explain strong fluorescence in this
system within a ''U-V'' model, we should therefore invoke solid state
effects. The red-shift due to interchain interactions could lower the
one-photon gap by a larger extent than the two-photon gap since the
lowest two-photon state is covalent in nature. This again implies that
the interchain interactions in these systems are much stronger than
in the weakly fluorescent polymers. The relative positions of these
lowest states in different symmetry subspaces could also shift if we
choose a larger $V$ in our ''U-V'' model. However, this would require 
an unphysically large $V$ for a given $U$. In substituted polymers, the
electron hole symmetry as well as the reflection symmetry are broken.
 This could lead to reasonable oscillator strengths to the lowest
singlet excitation resulting in strong fluorescence.

The ground state, the lowest one-photon,
two-photon and triplet states show very interesting bond-order
patterns. In Fig. (8), we present these data for the 8-ring
oligomer. In the ground state, the intra-ring bonds are very nearly
identical and the magnitude of the bond-order is close to that of 
an isolated benzene ring. The inter-ring bond is very weak and 
has a bond-order of $0.24$ and nearly the same for all the 
inter-ring bonds.

The bond-orders in the lowest one-photon state show very interesting
variations. Towards the ends of the oligomer, the rings are benzene
like while as we move to the interior, the rings acquire quinonoidal
characteristics. The inter-ring bonds  progressively strengthen
as we move towards the interior of the oligomer.
The bond-order pattern in lowest two-photon state is qualitatively
different from that of the $1 ^{1}B_{u}^{-}$ state. In this case
the rings in the exterior of the molecule are benzene like and the
inter-phenyl bonds in this region have very weak double bond 
character. In this region, the bond-orders resemble the bond-orders
in the ground state. However, in the interior ring, the intra-ring 
bond-orders
though nearly uniform have a value smaller than that in the exterior.
This shows that the excitation is mainly confined to the interior
two-rings in the full system. 
The lowest triplet state shows a strengthening of the inter-ring
bond as we move towards the center of the molecule. Furthermore,
the bonds involving the carbons at the para position become
weaker in the interior rings while those between the ortho and
meta carbons become stronger. This supports a quinonoidal structure
in the interior of the PPP system while away from the center
the rings tend to be more benzene like with the inter-ring bonds
having rather weak double bond character in this region.

The strong similarity in the bond-order patterns of the lowest 
triplet and the lowest $1^{1}B_{u}^{-}$ state allows us to visualize
the excitations as spread out over the entire oligomer with the
quinonoidal character strengthening as we move towards the interior.
However, the $2^{1}A_{g}^{+}$ state has a distinctly different
characteristic and appears to be a strongly localized excitation,
with the excitation essentially confined to two middle rings of the 8-ring
oligomer.
We would expect this difference between the $2^{1}A_{g}^{+}$ state,
the $1^{1}B_{u}^{-}$ state and the $1^{3}B_{u}^{+}$ states to increase
when the range of interactions are extended beyond the nearest
neighbour. This is because, the extended range interactions allow
for stability of structures in which charges are far apart. Such
resonance structures have very strong quinonoidal character. 
The local character of the $2^{1}A^{+}_{g}$ state shows that introducing
nonplanarity in the system will not shift the two-photon gap
significantly. Thus, we should expect the nonplaner PPP system to
fluroscece more weakly than the planer modification. 

The study of the doped states of the system is important
as it is conjectured that the excitations decay to give rise to charged 
segments. The photoinduced absorption corresponds to excitations
from these charged segments. We have shown in Fig. (6)
the energy per additional ring for one and two doped charges. We 
find that the finite size effect is more pronounced in doped systems
than in the undoped systems. However, the change in energy for
addition of a ring is almost the same as in the polaronic and bipolaronic
systems and is very close to the value in the undoped system. The DMRG
procedure conserves {\it electron hole} symmetry of the Hamiltonian as observed from
the difference in total energies between the oppositely charged species,
which corresponds to $nU$, where $n$ is the number of doped charges in the
system.

The bond orders of the positively and negatively charged polarons are
shown in Fig. (9). The positively charged polaron shows slightly 
pronounced quinonoidal character in the middle of the PPP system while
the negatively charged polaron shows more benzenoid character. The 
inter-ring bonds in both the cases have not acquired strong double bond
character in the interior, although the bonds in the interior ring of 
positively charged polaron shows quinonoidal characteristics.  The charge
density of the negatively charged polaron is almost smeared out uniformly
throughout  the system. However, the positively charged polaron shows 
(Fig. (10a))
slightly localized character in the interior rings. This is consistent
with the bond order differences between the two systems (Fig. (9)).  
The spin density in the negative polaron is spread out nearly uniformly at
equivalent sites (Fig. (10b)). The maximum difference in spin densities
between inequivalent sites is $\sim 0.11$, the highest positive spin
density is $0.06$ while the highest negative spin density is $-0.05$. In
the positive polaron the spin density variations are more towards
the ends of the system while for the negative polaron, it is more towards
the interior of the system, again consistent with the charge density 
and bond order data.

The positively charged and negatively charged bipolarons show almost
identical bond order patterns (Fig. (11)). The terminal rings are 
benzenoid in character while the interior rings have quinonoidal
character. This pattern is very similar to the bond order pattern in
the $1^{1}B_{u}^-$ and the $1^{3}B_u^+$ states. The charge density 
distributions (Fig. (12)) also show very similar behavior. The excess
charge is almost exclusively found at the ends of the system with the
interior rings being almost neutral.

It is possible to view the bipolarons,  $1^{1}B_u^-$ and $1^{3}B_u^+$
states as similar excitations. In the bipolaronic case we can 
look at
the excess charge as being confined to the ends of the chain and the 
system as being quinonoidal in the interior. The triplet state can be
viewed as the unpaired electrons being confined to the ends of the
system with the interior rings being quinonoidal. The $1^{1}B_u^-$
excited state can be viewed as an electron and a hole confined to the 
ends of the chain with the interior of the system again being 
quinonoidal. 

The most convincing evidence for the similarities amongst
bipolaronic ground state and $1^{1}B_u^-$ and $1^{3}B_u^+$ excited states 
of the neutral system comes from comparison of the appropriate correlation 
functions. 
The inter-site correlations we have studied are between the
 $24^{\prime}$ (the last site added in the DMRG procedure for the
eight ring system on the right-half, Fig. (13)), which is at the para-position
and the para-position sites of all the rings on the left ( shown in
upper panels of all the figures) on the one hand and the inter-site 
correlations
between site $24^{\prime}$ and the ortho- and meta-position sites 
of the left-half of the system on the other hand (bottom panels in all
the figures). The successive pairs of sites (e.g. 2 and 3,4 and 5, etc.)
are both ortho- or both meta- position sites
located  on the top and bottom half of the PPP system.

In Fig. (13a) we show the charge-charge correlation function for the 
dipole-allowed states ($1^{1}B_{u}^{-}$) and in Fig. (13b) is shown
the spin-spin correlation function  for the triplet state. There
is a dramatic similarity between the charge-charge correlation function
of the  $1^{1}B_u^-$ and spin-spin correlations of the $1^{3}B_u^+$ 
state. These excited states can be viewed as the charge and spin
counterparts of each other. The bipolarons also show strong charge-charge
correlations (Fig. (13c)) for sites between which charge-charge correlations 
of the $1^{1}B_u^-$ states are strong.  However, the excess charges in the 
bipolarons are more localized towards the ends of the system and 
hence the charge-charge correlations build up towards the exterior
of the system starting from the middle. In the 
$1^{1}B_u^-$ state, the correlation function dies down towards the 
exterior. If we were to deal within a periodic boundary condition,
this distribution would have vanished. This is indeed reflected in
the similarity  between the spin-spin correlation function (Fig. (13d))
of the $1^{3}B_u^+$ state and those of the bipolarons. The total picture
that emerges from the comparisons of spin-spin and charge-charge 
correlations of these four states shows that these states are indeed
very similar. Experimentally, this similarity should manifest as
comparable Stokes shifts in the fluorescence/phophorescence from
these states.

\section{Summary}
To conclude, we have carried out DMRG studies of the oligomers of
PPP system, within a ``U-V'' model. The parameters of the model are
determined by comparing the exact ``U-V'' model results with the exact
Pariser-Parr-Pople model results for biphenyl. The reliability
of the DMRG technique is illustrated by comparing the DMRG results for
$U=V=0$, with H\"{u}ckel results, for the largest system sizes studied.
In the ''U-V'' model, the one-photon state lies about 1eV above the 
two-photon state. Hence, to explain the strong fluorescence in this 
system, it is necessary to invoke interchain interactions which is 
expected to red-shift the one-photon state significantly more than 
the two-photon state. In  the PPP system, bond order
studies reveal that the $2^1A_g^+$ state corresponds to a localized excitation
while the $1^1B_u^-$ and the $1^3B_u^+$ excitations are extended in nature.
Furthermore, from the charge-charge and spin-spin correlation studies we 
find that the lowest one-photon state, the lowest triplet state and the 
bipolaronic ground state are very similar in nature and correspond to
 charge/spin separated 
configurations. The bipolaronic state shows clear indication of
disproportionation into two polarons in the polymeric chain.

{\bf Acknowledgment}: This work has been supported by the Indo-French
Centre for the Promotion of Advanced Research through project No. 1308-4,
"Chemistry and Physics of Molecule Based Materials".

\pagebreak
\begin{center}
{\bf Tables}
\end{center}

{\bf Table 1}: Comparison of excitation gaps of DMRG calculation (with cut-off
parameter $m=120$) with exact calculation for biphenyl.
The intersite electron-electron interaction parameter, $V$, is
4.5eV. All the energies are in eV, $\triangle E_{ST}$ corresponds to
lowest spin gap. The one-photon gap from exact Parisar-Parr-Pople
calculation is $4.8814$eV.   

\begin{center}
\begin{tabular}{|cccc|} \hline
method & one-photon gap & two-photon gap & $\triangle E_{ST}$ 
 \\ \hline
 Exact & 4.8490 & 3.3752 & 2.3665 \\
 DMRG & 4.8857 & 3.1301 & 2.4394  \\ \hline
\end{tabular}
\end{center}

{\bf Table 2}: Comparison of energies from DMRG calculations with cut-off $m=120$
with exact calculation for the noninteracting models of eight monomers 
of the PPP system. 

\begin{center}
\begin{tabular}{|ccc|} \hline 
doping & H\"{u}ckel  & DMRG  \\ \hline
0  & -158.518 & -158.664 \\
+1 &-157.272 & -156.996 \\
-1 &-157.272 & -156.432 \\
+2 &-156.026 & -155.394 \\
-2 &-156.026 & -155.334 \\ \hline
\end{tabular}
\end{center}

\pagebreak

\pagebreak

\begin{center}
{\bf Figure captions}
\end{center}

{\bf Fig.1}: Schematic diagram of the left half of Poly-para-phenylene (PPP)
system we have studied. The right half is related by $C_{2}$ symmetry.
The numbers in (a) and (b) correspond to the site indices and bond indices 
respectively. The middle bond is not shown in the figures.

{\bf Fig.2}: Schematic diagram of the building up of PPP oligomers by the DMRG
method.

{\bf Fig.3}: Comparison of bond orders for neutral PPP system corresponding to 
the longest system size we have studied, from H\"{u}ckel and
DMRG calculations. Squares represent exact and triangles represnt DMRG
calculations. The bond index is as shown in Fig. (1b).

{\bf Fig.4}: Comparison of bond orders for positive bipolaron of PPP system
corresponding to  the longest system size we have studied, from H\"{u}ckel and
DMRG calculations. Squares represent exact and triangles represnt DMRG
calculations.

{\bf Fig.5}: Comparison of bond orders for negative bipolaron of PPP system
corresponding to  the longest system size we have studied, from H\"{u}ckel and
DMRG calculations. Squares represent exact and triangles represnt DMRG
calculations.

{\bf Fig.6}: Change in energy (eV) per monomer vs. number of monomer units 
for PPP: neutral (squares),  polaronic (triangles) and  bipolaronic 
(diamonds) systems.

{\bf Fig.7}: Energy gaps (eV) for  $1^1A_g^+ \rightarrow 2^1A_g^+$ (squares),
$1^1A_g^+ \rightarrow 1^1B_u^-$ (triangles) 
and  $1^1A_g^+ \rightarrow 1^3B_u^+$ (diamonds) excitations for PPP
system.

{\bf Fig.8}: Bond orders for (a) $1^1A_g^+ $, (b) $2^1A_g^+ $,
(c) $1^1B_u^- $ and (d) $1^3B_u^+$  states of PPP.
Bond index corresponds to numbering given in Fig. (1b). 

{\bf Fig.9}: Bond orders  for (i) positively (squares) and (ii) negatively 
(triangles) charged polaron of PPP system. Bond index corresponds to the 
numbering given in Fig. (1b).

{\bf Fig.10}: (a) Charge density and (b) spin densities for singly doped
PPP system. In both (a) and (b), (i) corresponds to positively (squares) 
and (ii) to negatively (triangles) charged polarons.

{\bf Fig.11}: Bond orders  for (i) positively (squares) and (ii) negatively 
(triangles) charged bipolaron of PPP system. Bond index corresponds to 
the numbering given in Fig. (1b). 

{\bf Fig.12}: Charge densities  for (i) positively (squares) and (ii) negatively 
(triangles) charged bipolarons of PPP system. Site index corresponds to
the numbering given in Fig. (1a). 

{\bf Fig.13}: Charge-charge correlation functions of PPP system for 
(a) $1^1B_u^-$ state and (c) negative bipolaron and spin-spin correlation 
functions for (b) $1^3B_u^+$  and (d) negative bipolaron.
In all the four figures, upper panel corresponds to the correlation of
newly added site of the right block with  all the para-position sites 
of the left block while bottom panel corresponds to the correlation of
newly added site of the right block with all the ortho- and meta-position
sites of the left block. 
\end{document}